\documentclass[conference]{IEEEtran}
\usepackage{authblk,amsmath,amsfonts,amssymb,epsfig,color,hyperref}

\hypersetup{
    colorlinks,%
    citecolor=blue,%
    filecolor=blue,%
    linkcolor=blue,%
    urlcolor=blue
}
\definecolor{bluecolor}{rgb}{0,0.,1.}

\definecolor{redcolor}{rgb}{.7,0.,0.}

\newcommand{\pr}[1]{\left( #1\right)}
\newcommand{\prr}[1]{\left[ #1 \right]}
\newcommand{\es}[1]{\begin{equation}\begin{split}#1\end{split}\end{equation}}

\newcommand{\R}{\mathbb{R}}

\newcommand{\rr}{\mathbf{r}}
\newcommand{\dd}{\mathrm{d}}
\newcommand{\V}{\mathcal{V}}

\usepackage{fancyhdr}
\begin{document}

\title{Network Connectivity: Stochastic vs. Deterministic Wireless Channels}
\author[1,2]{Orestis Georgiou\thanks{orestis.georgiou@toshiba-trel.com}}
\author[2]{Carl P. Dettmann}
\author[3]{Justin P. Coon}
\affil[1]{Toshiba Telecommunications Research Laboratory, 32 Queens Square, Bristol, BS1 4ND, UK}
\affil[2]{School of Mathematics, University of Bristol, University Walk, Bristol, BS8 1TW, UK}
\affil[3]{Department of Engineering Science, University of Oxford, Parks Road, Oxford, OX1 3PJ, UK}
\maketitle

\pagestyle{plain}

\thispagestyle{fancy}

\begin{abstract}
We study the effect of stochastic wireless channel models on the connectivity of ad hoc networks.
Unlike in the deterministic geometric disk model where nodes connect if they are within a certain distance from each other, stochastic models attempt to capture small-scale fading effects due to shadowing and multipath received signals.
Through analysis of local and global network observables, we present conclusive evidence suggesting that network behaviour is highly dependent upon whether a stochastic or deterministic connection model is employed.
Specifically we show that the network mean degree is lower (higher) for stochastic wireless channels than for deterministic ones, if the path loss exponent is greater (lesser) than the spatial dimension.
Similarly, the probability of forming isolated pairs of nodes in an otherwise dense random network is much less for stochastic wireless channels than for deterministic ones.
The latter realisation explains why the upper bound of $k$-connectivity is tighter for stochastic wireless channels.
We obtain closed form analytic results and compare to extensive numerical simulations.
%
\end{abstract}



\section{Introduction \label{sec:intro}}

Recent advancements in micro and nano-scale electronics along with the development of efficient routing protocols have rendered current wireless technologies ideal for ad hoc and sensing applications \cite{cordeiro2011ad}.
Making use of low complexity multi-hop relaying techniques and signal processing capabilities, sensor networks can often achieve very good coverage and connectivity over large areas, ``on the fly'' in a decentralized and distributed manner by self-organising into a mesh network, assigned with some data collection and dissemination task \cite{savazzi2013ultra}.

The spontaneous self-organization trait of large scale sensor networks has attracted much research attention in recent years, with particular interest in enhancing the connectivity properties of the underlying communication graph \cite{santi2005topology}.
Of practical interest for example is the ability to predict the optimal number of nodes necessary to maintain good connectivity \cite{xue2004number}, or conversely, to predict the optimal average transmission range for a given number of nodes.
These predictions are essential network design recommendations which can in turn act as inputs for cognitive scheduling and routing protocol selection.
Furthermore, improvement in the connectivity properties of the network can have a significant impact on the operational lifetime of ad hoc sensor networks by conserving energy, and can also improve the overall functionality, reliability and fault-tolerance of the network.

\begin{figure}[t]
\begin{center}
\includegraphics[scale=0.2]{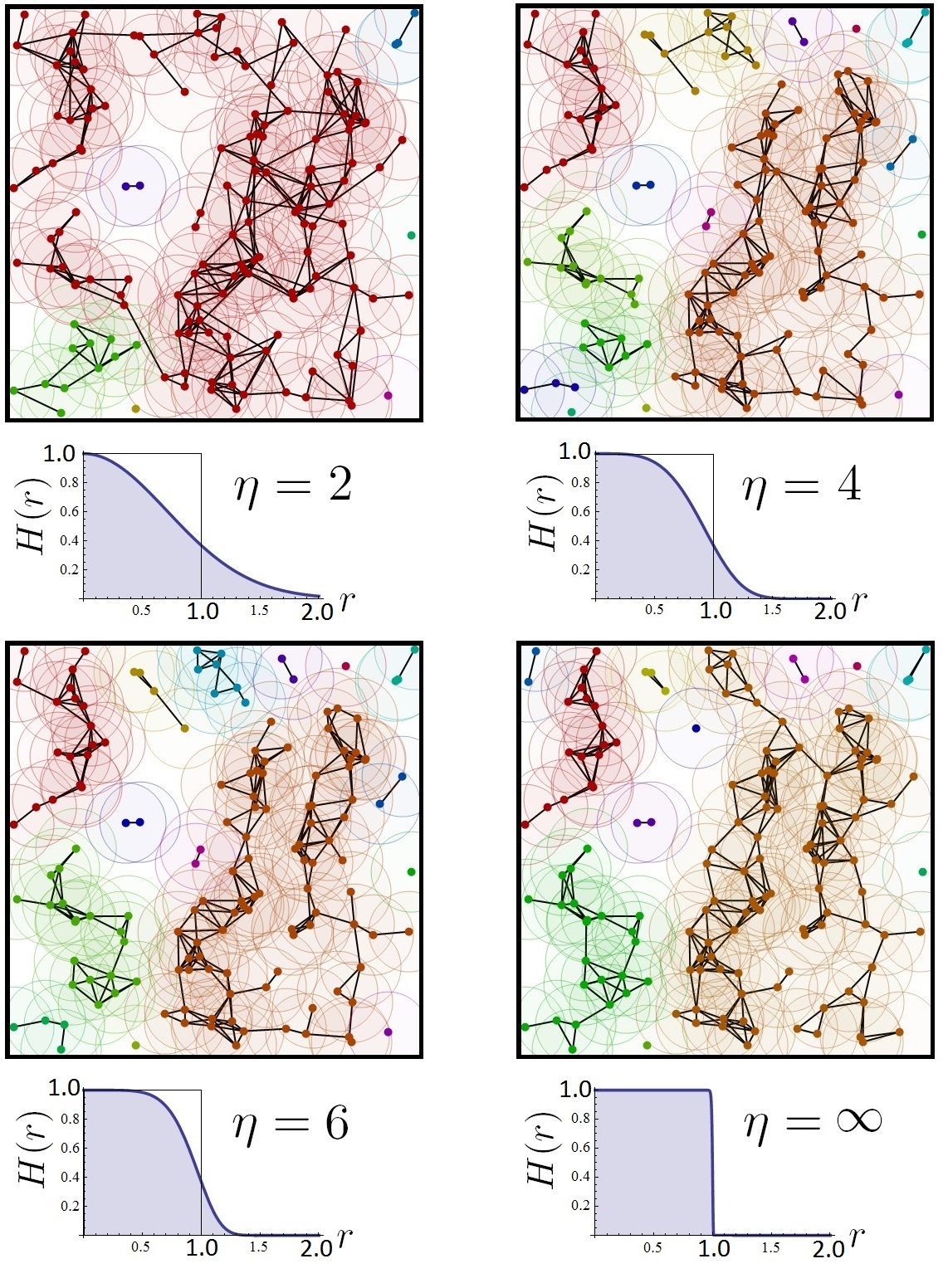}
\caption{\label{fig:real} 
Random realizations of ad hoc networks in a square domain of side $L=10$, using $\beta=1$ and $\eta=2,4,6,\infty$, as defined in the pair connectedness function $H(r)$.
Different connected components are shown in different colors.}
\end{center}
\end{figure}

The analysis and resolution of network connectivity is typically addressed from a physical layer's point of view through the theory of stochastic geometry \cite{stoyan1995stochastic}, random geometric graphs \cite{penrose2003random} and complex networks \cite{albert2002statistical},
equipped with a plethora of methods and metrics e.g. clustering and modularity statistics, node importance, correlations between degrees of neighbouring nodes, \textit{etcetera}. 
From a communications perspective, a popular and well studied observable of random networks is that of \textit{full connectivity} \cite{coon2012full}. 
This characterizes the probability $P_{fc}$ with which a random realization of an ad hoc network will consist of a single connected component (cluster). 
Consequently, every node in the network can communicate with any other node in a multi-hop fashion.

In the theory of random geometric networks, two nodes are said to connect and form a pair if they are within a certain distance $r_0$ from each other. 
This is referred to as the \textit{geometric disk model} and is only sufficient when modelling deterministic, distance-dependent wireless channels.
One extension of this connectivity model, is to account for the channel randomness due to shadow fading and multipath effects in the spatial domain by `softening' the position dependence of the pair connectedness function such that links are established in probability space \cite{bettstetter2005connectivity}.
In static wireless networks, this softening can be understood as modelling the randomness in the received signal power \cite{miorandi2008impact}.
Thus, in the absence of inter-node interference, two nodes connect with a probability $H(r)$ which decays as a function of the distance $r$ between the pair.
What this effectively means is that nodes which are closer than a distance $r_0$ from each other may no longer be connected, and nodes that are more than $r_0$ apart may be connected.

The overall effect of channel randomness is typically encoded into the path loss exponent $\eta$ (defined later) \cite{tse2005fundamentals} and has been argued to have a negative effect on \textit{local} connectivity, but a positive effect on \textit{overall} connectivity \cite{zhou2008connectivity}.
The latter conclusion has been reached through a combination of theoretical and numerical results, the former mainly relying on the bound $P_{fc}\leq P_{md}$, where $P_{md}$ characterizes the probability with which a random realization of an ad hoc network has minimum network degree equal to one i.e. each node is connected to at least one other node.
The bound was proven to be tight by Penrose in 1999 for the deterministic geometric disk model in the limit of number of nodes $N\to\infty$ \cite{penrose1999k}.
Since then, a number of papers have followed suit extending these results in many directions to include for, \textit{inter alia}, boundary effects, channel randomness, and anisotropic radiation patterns.

In this paper, we challenge the negative effect of channel randomness on \textit{local} connectivity and also investigate the tightness of the bound $P_{fc}\leq P_{md}$ for finite yet sufficiently large $N$ and examine the role of channel randomness to this respect. 
We show analytically that \textit{Rayleigh fading} improves local connectivity when $\eta$ is less than the effective spatial dimension $d$ of the network.
We also show that the \textit{pair isolation probability} $\Pi$ is what distinguishes $P_{md}$ from $P_{fc}$ at high node densities, and calculate closed form expressions for it assuming a Rayleigh fading channel. 
Both of our results are validated through extensive numerical simulations, the latter suggesting that two nodes are more likely to form an isolated pair when $\eta$ is large i.e. in heavily cluttered environments.
Finally, we discuss the engineering insight provided by our analysis towards
facilitating the design of wireless ad hoc sensor and mesh networks \cite{haenggi2009stochastic}.

The paper is structured as follows: 
Sec.~\ref{sec:setup} describes the system model and relevant assumptions. 
Secs.~\ref{sec:MD} and~\ref{sec:connectivity} discuss local and global network observables respectively and their sensitivity to stochastic/deterministic wireless channels. 
Sec.~\ref{sec:TA} investigates analytically the pair isolation probability, and
Sec.~\ref{sec:NS} numerically confirms our theoretical predictions which are then summarised and discussed in Sec.~\ref{sec:CD}.


\section{System Model \label{sec:setup}}

We consider a network of $N$ nodes distributed randomly and uniformly in a $d=2,3$, dimensional convex domain $\V \subset\R^d$ with volume $V$.
The node position coordinates are given by $\rr_i \in \V$ for $i=1\ldots N$.
We say that a communication link between a pair of nodes $i$ and $j$ exists with probability $H(r_{ij})$, where $r_{ij}= |\rr_i - \rr_j|$ is the relative distance between the pair.
One physical interpretation of a communication link is given by the complement of the information outage probability between two nodes for a given rate
$x$ in bits per complex dimension, which can be written as
$\textrm{Pr}\pr{\log_2 (1+ \textrm{SNR} \times |h|^2) > x}$ \cite{tse2005fundamentals}, 
where $h$ is the channel transfer coefficient, and $\textrm{SNR} \propto r_{ij}^{-\eta}$ is the average received signal to noise ratio and $\eta$ is the path loss exponent. Typically $\eta=2$ corresponds to propagation in free space but for cluttered environments it is observed to be $\eta \geq 2$.

We consider the case where the individual channel fading distributions follow a Rayleigh fading model,
and all channels are statistically independent. 
It follows that $|h|^2$ has a standard exponential distribution for a single-input single-output (SISO) antenna system.
Hence, the connection probability $H(r)$ between two nodes a distance $r$ apart can be expressed as
\es{
H(r)= e^{-\beta r^{\eta}}
\label{H},}
where $\beta$ sets the characteristic connection length $r_0=\beta^{-1/\eta}$.
Therefore, our system model has two sources of randomness: random node positions, and random link formation according to the `softness' of the channel fading model controlled here by $\eta$.
It is important to note that in the limit of $\eta\rightarrow \infty$, the connection between nodes is no longer probabilistic and converges to the geometric disk model, with an \textit{on/off} connection range at the limiting $r_0$.
We will later make use of this limit in order to compare the connectivity of random networks using deterministic or stochastic point-to-point link models.

Fig.~\ref{fig:real} shows how $N=150$ nodes scattered randomly in a square domain of side $L=10$, connect to form different networks for different values of $\eta=2,4,6,$ and $\infty$, using $\beta=1$.
The corresponding $H(r)$ functions are also plotted below each panel in order to give the reader a feeling of how probable shorter and longer than $r_0$ links are in the presence of Rayleigh fading. 
We stress that in practice, an efficient medium access control (MAC) layer protocol is typically required e.g. using a Time Division Multiple Access (TDMA) scheme in order to render inter-node interference negligible. Alternatively, a low traffic network may be assumed so that the communication network can be modelled as seen in Fig.~\ref{fig:real}.


\section{Local Connectivity and Mean Degree \label{sec:MD}}

The network mean degree $\mu$ is a local observable of network connectivity characterising the average number of one-hop neighbours of a \textit{typical} node. 
For random geometric networks as described in Sec.~\ref{sec:setup}, this can be expressed as $\mu= \frac{N-1}{V^2}\int_{\V^2} H(r_{ij}) \dd \rr_i \dd \rr_j
$ which follows from multiplying $N-1$ by the probability of two randomly selected nodes ($i$ and $j$) connecting to form a pair.
Assuming that $V$ is large and the typical length scale of the domain is greater than the effective connection range $r_0$, a typical node is most likely to be found away from the borders of the domain.
Moreover, since $H(r)$ is decaying exponentially, it is reasonable to expect that the degree of a typical node is very much insensitive to boundary effects, thus justifying the following approximation
\es{
\mu_\eta &\approx \rho \int_{\R^{d}} \!\! H(r) \dd \rr = \rho \, \Omega \int_0^\infty \!\! r^{d-1} e^{-\beta r^{\eta}} \dd r = \rho \frac{\Omega \Gamma(\frac{d}{\eta})}{\eta \beta^{\frac{d}{\eta}}}
\label{mu}}
where $\rho\approx\! \frac{N-1}{V}$ and $\Omega=\!\frac{2\pi^{\frac{d}{2}}}{\Gamma(\frac{d}{2})}$ is the solid angle in $d$ dimensions.
In \eqref{mu} we have effectively assumed that the network is homogeneous i.e. it is translation and rotation invariant, such that we can set $\rr_i=\mathbf{0}$ and extend the radial integral of $\rr_j$ to infinity taking on exponentially small errors.
Substituting $\beta=r_0^{-\eta}$ and taking the limit $\eta \to\infty$ we obtain an expression for the mean degree for the deterministic geometric disk model $\mu_\infty = \rho \, \Omega \frac{r_0^d}{d}$.
Comparing this to the result of \eqref{mu}, we conclude that $\mu_\infty = \mu_\eta$ when $\eta=d$ and that $\mu_\eta< \mu_\infty$ for $\eta>d$.
In other words, the stochastic and deterministic connectivity models are equivalent at $\eta=d$ when viewed locally.
More importantly however, Rayleigh fading reduces the local network connectivity when $\eta>d$, but improves it when $\eta<d$.
The latter condition describes a very special case when the network resides in an effectively 2D plane, but may well be significant in 3D networks deployed in multi-storey buildings for example where the path loss exponent may be in the range $2\leq \eta <3$.
Consequently, one may significantly over or under specify network design features and deployment methods if the network is not correctly modelled.

\begin{figure}[t]
\begin{center}
\includegraphics[scale=0.3]{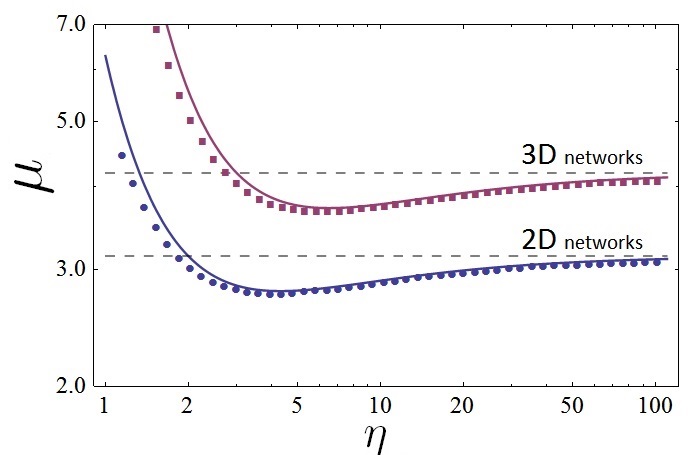}
\caption{\label{fig:mean} 
Plot of the mean degree $\mu$ as a function of $\eta$, obtained from numerical simulations of ad hoc networks in a square of side $L=10$ (blue), and a cube of side $L=7$ (purple). In both cases $\beta=\rho=1$. The analytical predictions of \eqref{mu} are shown as solid curves and the limit $\mu_\infty$ as dashed lines.
}
\end{center}
\end{figure}

Numerical verification of the above result is presented in Fig.~\ref{fig:mean} showing computer Monte Carlo simulations of ad hoc networks in two and three dimensions. 
The analytical predication of \eqref{mu} and the limit case of $\mu_\infty$ are also shown for comparison.
An almost perfect agreement is observed between theory and simulations with the theoretical result typically being smaller than the numerical one.
This is because $\mu_\eta$ was approximated assuming no boundary effects, which hinder connectivity for nodes near the borders of $\V$.
The minimum value of $\mu_\eta$ is obtained numerically at $\eta\approx4.33$ and $\eta\approx6.50$ in two and three dimensional networks respectively.
We now turn to investigate global network observables.


\section{$k$-Connectivity and Minimum Degree \label{sec:connectivity}}

In the absence of a fixed infrastructure (e.g. cellular, or WLANs), where it is sufficient that each network node has a wireless link to at least one access point, in decentralized ad hoc networks, efficient routing protocols can utilize mutually independent paths to communicate information through the network e.g. for sensing, monitoring, alerting or storage purposes \cite{cordeiro2011ad}.
Therefore, if a multihop path exists between all pairs of nodes, then the network is fully connected and in a sense is both delay and disruption tolerant \cite{fall2008dtn}.

A generalization of the concept of full connectivity is that of $k-$connectivity \cite{georgiou2013k}.
A fully connected network is said to be $k$-connected if the removal of any $k-1$ nodes leaves the remaining network fully connected.
The removal of nodes may model technical failures (e.g. a hardware/software malfunction) or attacks which can disrupt the functionality and operation of the network, in some cases leading to cascades of catastrophic failures \cite{buldyrev2010catastrophic}.
Equivalently, $k$-connectivity also guarantees that for each pair of nodes there exist at least $k$ mutually independent paths connecting them \cite{penrose2003random}.
Therefore, $k$-connectivity is an important measure of network robustness, resilience but also of routing diversity.

It is clear that a $k$-connected network has minimum degree $k$, i.e. each node has at least $k$ neighbouring nodes. 
The opposite statement is not true however and hence the former set is a subset of the latter and so $P_{fc}(k)\leq P_{md}(k)$, where $P_{fc}(k)$ denotes the probability that a random realization of an ad hoc network is $k$-connected, and $P_{md}(k)$ denotes the probability that it has minimum degree $k$. 
These two observables are however strongly related through a fundamental concept originally proven in \cite{penrose1999k} (Theorem 1.1) which states that ``if $N$ is big enough, then with high probability, if one starts with isolated points and then adds edges connecting the points in order of increasing separation length, then the resulting graph becomes $k$-connected at the instant when it achieves a minimum degree of $k$''.
Ever since this realization, $P_{fc}(k)$ has been approximated by $P_{md}(k)$ \cite{bettstetter2004connectivity} as it is easier to express mathematically, and evaluate numerically.
Specifically, we have that \cite{georgiou2013k}
\es{
P_{md}(k)&= \langle\prod_{i=1}^{N} P(\textrm{degree}(\textbf{r}_{i})\geq k) \rangle \\
&\approx \!\prr{1-\! \sum_{m=0}^{k-1}\!\frac{\rho^{m}}{m!} \frac{1}{V}\int_{\mathcal{V}}\!\! M_{H}^{m} (\textbf{r}_i) e^{-\rho M_{H} (\textbf{r}_i)}
\textrm{d}  \textbf{r}_i}^{N}
,
\label{Pmd}}
where $M_{H}(\rr_i)\!\!= \!\! \int_{\V} H(r_{ij}) \dd \rr_j$ and $V$ is assumed to be much larger than $\pi r_0^d$.
The angled brackets in \eqref{Pmd} represent a spatial average of a network observable $O$ over all possible node configurations and is defined as
\es{
\langle O\rangle = \frac{1}{V^{N}}\int_{\mathcal{V}^{N}} O(\mathbf{r}_{1},\mathbf{r}_{2},\ldots, \mathbf{r}_{N}) \mathrm{ d}\mathbf{r}_{1} \mathrm{ d}\mathbf{r}_{2}\ldots \mathrm{ d}\mathbf{r}_{N}
\label{average}
.}
We will use this notation in order to calculate expectation values of different random network observables.

\begin{figure}[t]
\begin{center}
\includegraphics[scale=0.2]{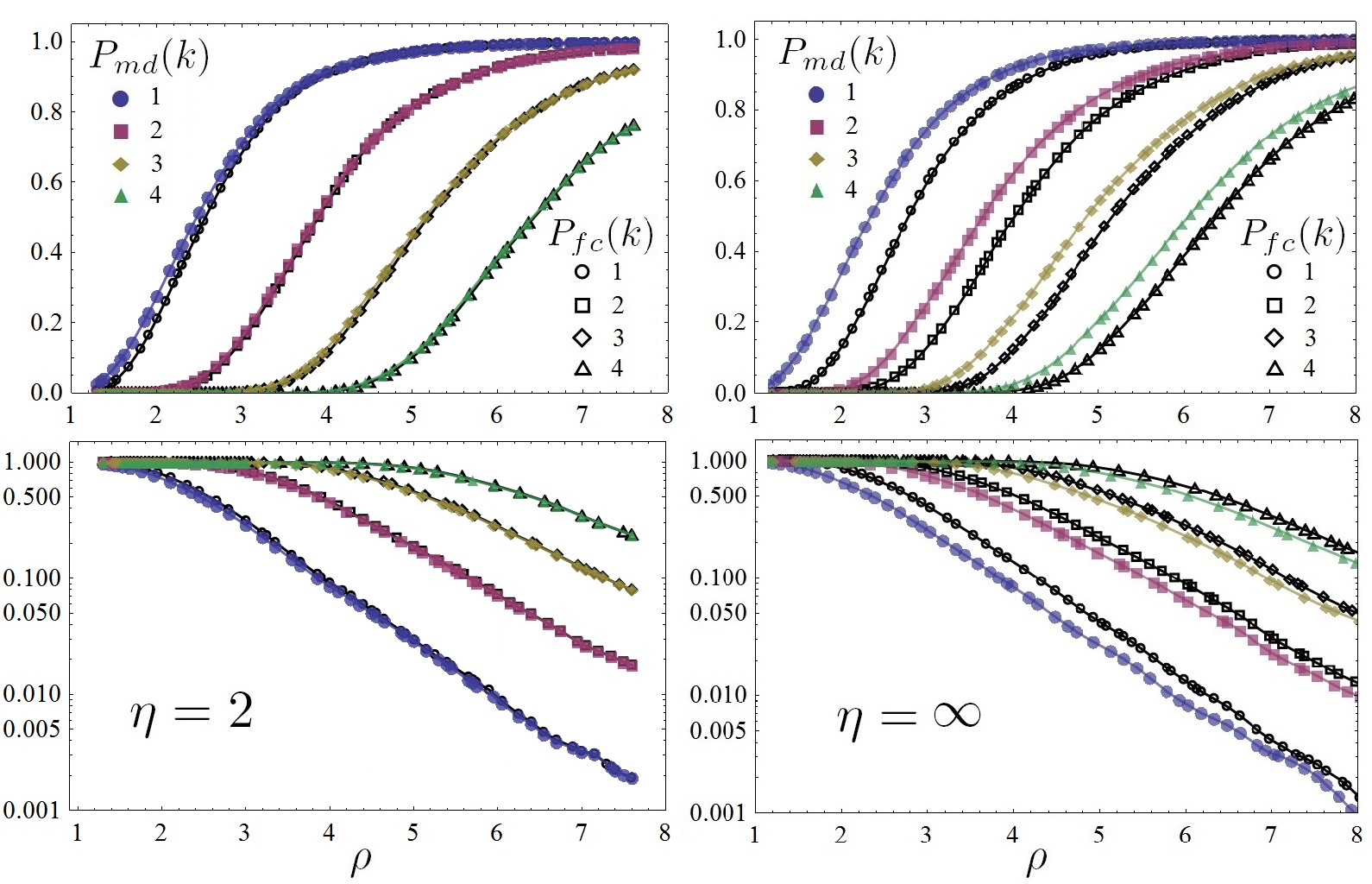}
\caption{\label{fig:soft} 
Computer simulation of $P_{md}(k)$ (filled markers) and $P_{fc}(k)$ (hollow markers) for $k\in[1,4]$ using $\beta=1$ and $\eta=2$ (\textit{left}) and $\eta=\infty$ (\textit{right}) in a square-shaped domain of side length $L=10$. \textit{Bottom:} $1-P_{md}(k)$ and $1-P_{fc}(k)$ on a log-linear scale. }
\end{center}
\end{figure}

Fig.~\ref{fig:soft} shows Monte Carlo computer simulations for $\eta=2$ (left panels) and $\eta=\infty$ (right panels). 
$10^5$ random networks were simulated in a square domain of side $L=10$ for a range of node density values $\rho\in(1,8)$ using $\beta=1$, in order to obtain curves for $P_{fc}(k)$ and $P_{md}(k)$ for $k=1,2,3,$ and $4$.
Indeed, it can be seen from Fig.~\ref{fig:soft} that the two observables are in good agreement with each other, with $P_{fc}(k)$ following $P_{md}(k)$ closely from below.
What is also evident however, is that the gap between the two $\Delta(k)=P_{md}(k) - P_{fc}(k)$ is significantly larger and persistent even at high densities for deterministic ($\eta=\infty$) rather than probabilistic ($\eta<\infty$) wireless channels.
The visible difference can be seen to persist even when plotted on a log-linear scale (bottom panels).
This observation has been investigated numerically in several articles \cite{bettstetter2005connectivity,miorandi2008impact,zhou2008connectivity,bettstetter2004connectivity} but has not been fully understood.
What is the impact of channel randomness?
Is fading good or bad?
These are but a few related questions we aim to revisit in the next section by attempting to understand where $\Delta(k)$ stems from and the significance of the path loss exponent $\eta$.


\section{Theoretical Analysis \label{sec:TA}}

It was recently shown in \cite{coon2012full}, that $P_{fc}(1)$ at high node densities is given by the complement of the probability of an isolated node
\es{
P_{fc}(1)= 1- \rho \int_\V e^{-\rho \int_\V H(r_{ij}) \dd \rr_j} \dd \rr_i + \ldots
,\label{pfc}}
where the $(\ldots)$ indicate higher order terms, possibly exponentially smaller than the leading order term.
Physically, \eqref{pfc} makes sense since a dense network which is not fully connected will most probably involve a single isolated node.
It is thus reasonable to extend this argument and expect that at high node densities a network which has minimum degree $1$ but is not fully connected will most likely consist of a large $N-2$ cluster and an isolated \textit{pair} of nodes. 
We therefore define $\Pi(1)$ as the probability that two randomly selected nodes are connected to each other, but are isolated from the remaining nodes of the network, and argue that $\Delta(1)=P_{md}(1) - P_{fc}(1) \approx \Pi(1)$ at high node densities. 
In order to confirm our hypothesis, we will analytically calculate the dominant contribution to $\Pi(1)$ and then compare against numerical simulations of $\Delta(1)$ for different values of $\eta$ obtained.

Using the spatial average defined in \eqref{average}, and letting $H_{ij} = H(r_{ij})$ in order to save space, we write 
\es{
\Pi(1)&= \langle \sum_{i<j} H_{ij} \prod_{k\not=j\not=i} (1-H_{ik})(1-H_{jk}) \rangle \\
&= \frac{N(N-1)}{2}\langle H_{ij} \prod_{k\not=j\not=i} (1-H_{ik})(1-H_{jk}) \rangle \\
&\approx  \frac{\rho^2}{2} \!\! \int_{\V^2} \! H_{ij} \! \prr{ \frac{1}{V} \! \int_{\V} (1-H_{ik})(1-H_{jk}) \dd \rr_k }^{N-2} \!\!\! \dd \rr_i \dd \rr_j \\
&\approx \frac{\rho^2}{2} \int_{\V^2} H_{ij}  e^{-\rho \int_{\V} H_{ik} + H_{jk} - H_{ik}H_{jk}  \dd \rr_k }
 \dd \rr_i \dd \rr_j
\label{delta}
,}
where we have assumed that $N\gg1$ such that $(N-1)/V \approx (N-2)/V \approx \rho$, and $(1-x)^{N} \approx e^{-N x}$.
In the third line of \eqref{delta}, we used the fact that $N-2$ of the $N$ integrals defined in \eqref{average} are separable since the possible connections of the pair $(i,j)$ to the remaining $N-2$ nodes are statistically independent events.
Whilst equation \eqref{delta} appears to be very complicated, there are several important observations to be highlighted which can simplify it. 
Firstly, in the high density limit we expect equation \eqref{delta} to be dominated by contributions where the inner integral in the exponent is small.
The integral in the exponent represents the probability that a randomly selected node $k$ connects with node $i$ or node $j$ or both, an event least probable if both $i$ and $j$ are near the boundary of the domain which physically corresponds to the most \textit{hard to connect to} region of $\mathcal{V}$.
Secondly, the dominant contribution of \eqref{delta} also requires $H_{ij} \approx 1$, and thus nodes $i$ and $j$ must be close to each other in order to form the isolated pair. 
Both these conditions are met at a corner of $\V$.

We therefore consider each of the four right angled corners of a square\footnote{Note that the current analysis is not restricted to the simple case of a square domain, nor the SISO point-to-point Rayleigh fading model \eqref{H}.} domain $\V=[0,L]^2$ independently and attempt to calculate $\Pi(1)$ from \eqref{delta}.  
We assume that nodes $i$ and $j$ are both near this corner and hence Taylor expand $H_{ik}$ and $H_{jk}$ in the integrand of in the exponential using polar coordinates up to linear order in $r_i$ and $r_j$ respectively to obtain
\es{ \hat{K}(\textbf{r}_i, \textbf{r}_j)&=
\int_{\V} H_{ik} + H_{jk} - H_{ik} H_{jk}\textrm{d} \textbf{r}_{k} \\
&\approx 
2^{-\frac{\eta+2}{\eta}}\beta^{-\frac{2}{\eta}}\Big[
\frac{\pi}{2}(2^{\frac{\eta+2}{\eta}}-1) \Gamma\pr{\frac{\eta+2}{\eta}}\\
&+ (2\beta)^{\frac{1}{\eta}} (2^{\frac{\eta+1}{\eta}}-1)
\Gamma\pr{\frac{\eta+1}{\eta}} \\
&\times \pr{ r_i (\cos \theta_i +\sin \theta_i) + r_j (\cos\theta_j +\sin \theta_j) }
\Big]
,
\label{ij}}
where we have ignored correction terms of order $\mathcal{O}(r_i r_j)$.
The integration limits in \eqref{ij} were taken to be $r_k\in[0,\infty)$ and $\theta_k \in[0,\pi/2]$.
The semi-infinite integration is allowed here since $H(r)$ decays exponentially fast with $r$ and so if $\beta L\gg 1$ any added errors in \eqref{ij} will be exponentially small.

\begin{figure}[t]
\begin{center}
\includegraphics[scale=0.25]{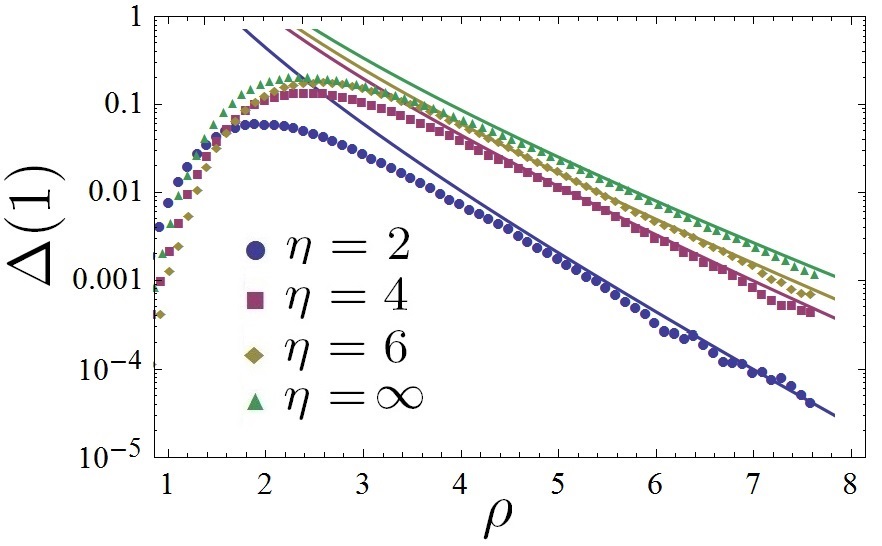}
\caption{\label{fig:delta}  
Computer simulations (filled markers) in a square domain showing $\Delta(1)=P_{md}(1)-P_{fc}(1)$ as a function of the density $\rho$ for different values of $\eta=2,4,6, \infty$ and $\beta=1$.  
The analytical approximation of $\Pi(1)$ is also plotted (solid curves).
}
\end{center}
\end{figure}

Substituting back into \eqref{delta} and assuming that the two nodes are sufficiently close such that $H_{ij} \approx 1$, we may now perform the remaining integrals to arrive at our main result
\es{
\Pi(1) \approx 32 \frac{2^{\frac{4}{\eta}} \beta^{\frac{4}{\eta}} \exp\!\prr{ -\rho \frac{\pi}{2} (1-2^{-\frac{\eta+2}{\eta}})\beta^{-\frac{2}{\eta}} \Gamma\pr{\frac{\eta+2}{\eta}}  }}
{\rho^2 \pr{(2^{-\frac{\eta+1}{\eta}}-1) \Gamma\pr{\frac{\eta+1}{\eta}} }^4  }
,
\label{PI}}
where we have also multiplied the result by $4$ due to the four identical right angle corners of $\V$. 
The resulting approximation of $\Pi(1)$ is a monotonically decreasing (increasing) function with respect to the density $\rho$ (path loss exponent $\eta$), which diverges like $\rho^{-2}$ near the origin (see Fig.~\ref{fig:delta}).
Therefore, in order for \eqref{PI} to be interpreted as a probability, one needs to carefully investigate regions of convergence and bound any approximating errors.
We will refrain from doing so for the sake of brevity and instead aim at obtaining useful engineering insight on the impact of stochastic wireless channel models.
We therefore substitute $\beta=r_0^{-\eta}$ into \eqref{PI} and expand for $\eta\gg1$ 
\es{
\Pi(1) \approx 32 \frac{ \exp\prr{-\rho \frac{\pi r_0^2}{4} + f(\eta) }}{r_0^{4}\rho^2}  +\mathcal{O}(1/\eta)
,
\label{asy}}
where $f(\eta)$ is always negative and decays monotonically to zero like $|f(\eta)|\sim\eta^{-1}$.
Note that the $\pi r_0^2 /4$ term appearing in the exponential indicates the available connection region for the pair to establish a link with the rest of the network.
We can thus conclude that the probability that two nearby nodes form an isolated pair is much less for stochastic $H(r)$ than for deterministic ones.
Moreover, comparing \eqref{asy} to the probability of an isolated \textit{node} in a square domain given by \cite{coon2012full}
\es{
4\frac{\exp\prr{-\rho \frac{\pi r_0^2}{4}}}{r_0^2 \rho}
\label{prev}
,}
in the high density limit, we observe that for a stochastic connectivity function (i.e. $\eta<\infty$), $\Pi(1)$ is exponentially smaller than \eqref{prev} whilst for a deterministic connectivity model (i.e. $\eta=\infty$) it differs only by an algebraic function of $\rho$. 

We have shown that the probability that two nearby nodes form an isolated pair is much less for stochastic $H(r)$ than for deterministic ones in the dense regime.
Furthermore, there is good reason to expect that this observation is not only restricted to Rayleigh fading channels but also to other small-scale fading models such as for example the \textit{two-wave with diffuse power} (TWDP) model \cite{durgin2002new}.
We postpone further discussion on the physical interpretation of this observation to Sec.~\ref{sec:CD} and turn to numerical simulations in order to verify that $\Delta(1) \approx \Pi(1)$ at high node densities.


\section{Numerical Simulations \label{sec:NS}}

In this section we will numerically investigate $\Delta(1)=P_{md}(1) - P_{fc}(1)$ obtained from computer Monte Carlo simulations of ad hoc networks (as in Fig.~\ref{fig:soft}) for different values of $\eta=2,4,6,\infty$, in a square domain of side $L=10$.
Our results are plotted in Fig.~\ref{fig:delta} and are contrasted against the theoretical prediction of equations \eqref{PI} and \eqref{asy}.
We observe that $\Delta(1)$ is a unimodal function of $\rho$, which peaks around $\rho\in(2,3)$ with a value of approximately $0.1$. 
Therefore, at such densities, approximately $1$ out of $10$ random realisations of ad hoc networks have minimum degree one, but are not fully connected.
At lower densities $\rho<2$, the difference $\Delta(1)$ is small because the set space of possible graphs is small, essentially enforcing a large overlap between full connectivity and minimum degree.
At higher node densities, where the set space grows exponentially, the difference between the two observables grows to a maximum, only to decay exponentially as the two distributions converge to $1$.
Fig.~\ref{fig:delta} confirms with a surprisingly good accuracy that $\Delta(1) \approx \Pi(1)$ at high node densities, and therefore a network which has minimum degree $1$ but is not fully connected will most likely consist of a large $N-2$ cluster and an isolated \textit{pair} of nodes. 
Moreover, the probability that two nearby nodes form an isolated pair is much less for stochastic $H(r)$ than for deterministic ones.

\begin{figure}[t]
\begin{center}
\includegraphics[scale=0.35]{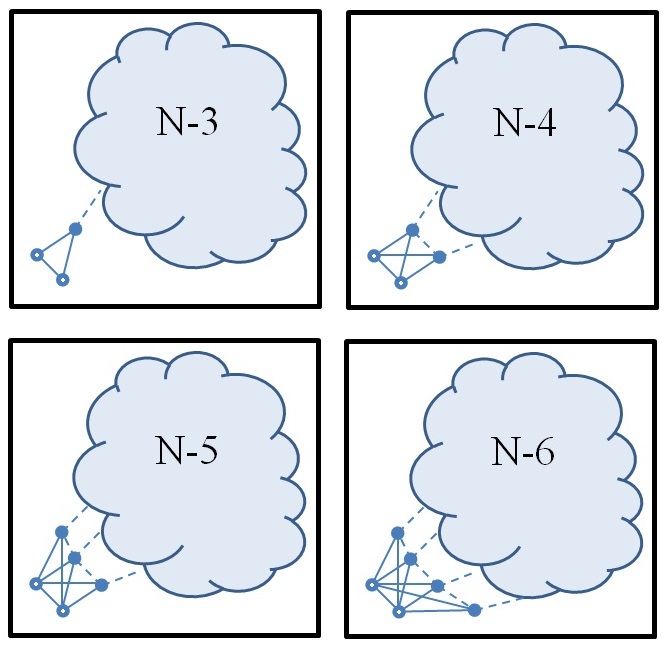}
\caption{\label{fig:egs} 
Example illustrations of networks involving a $k-1$ connected network of minimum degree $k$ for $k=2,3,4,5$. 
The clouds in each case are assumed to be $k$-connected networks of $N-k-1$ nodes, linked to a pair of nodes (which most probably is situated near a corner of $\V$) through $k-1$ ``bridging'' nodes. Note that for $k=1$, there is no bridging node.
}
\end{center}
\end{figure}

Given the numerical verification of our hypothesis, we may speculate for the more general case $k\geq1$, and conjecture that the dominant contribution to $\Delta(k)$ can be derived by considering the most likely scenario involving a $(k-1)$-connected network of minimum degree $k$ in the high density limit.
For example, for $k=2$ this would correspond to a pair of nodes which connect to the main cluster via a single node.
Similarly, for $k=3$ this would correspond to a pair of nodes which connect to the main cluster via 2 nodes.
See Fig.~\ref{fig:egs} for more illustrative examples for $k=2,3,4,5$.
We may therefore attempt to write down a general expression for $\Pi(k) \approx \Delta(k)$
\es{
\Pi(k)=\langle 
\sum_{i<j} H_{ij} \!\! \prod_{\substack{n=1\\ t_n\not=i,j}}^{k-1} \!\! H_{i t_n} H_{j t_n} \!\! \prod_{m\not=i,j,t_n} \!\!\! (1-H_{i m}) (1-H_{j m})
\rangle
\label{gen}
,}
where the nodes $t_n$ with $n<k$ act as relays or ``bridging'' nodes, which if removed, would cut off the pair $(i,j)$.

We will not however investigate $\eqref{gen}$ any further as it is beyond the scope of this paper.
Instead, we discuss briefly the concept of $k$-edge-connectivity related to the removal/failure of $k-1$ \textit{edges} (rather than nodes).
Clearly, a $k$-connected network is $k$-edge-connected and so $P_{fc}^{(n)}(k) \leq P_{fc}^{(e)}(k)$, where we have used superscript $(n)$ and $(e)$ to denote node-connectivity and edge-connectivity respectively.
This is because when you remove a node from a connected network, you always remove at least one edge.
Further, it is also clear that $P_{fc}^{(e)}(k) \leq P_{md}^{(n)}(k)$ since the former implies the latter and so we have that $P_{fc}^{(e)}(k)$ is \textit{sandwiched} between the two.
We can squeeze this bound further by noting that $k$-edge-connectivity implies the examples illustrated in Fig.~\ref{fig:egs} so that $P_{fc}^{(n)}(k) \leq P_{fc}^{(e)}(k) \leq P_{fc}^{(n)}(k) + \Pi(k) \leq P_{md}^{(n)}(k)$ as $\rho\to\infty$.
Such bounds are of interest to wireless network system designers, researchers as well as to the extended random graph community.


\section{Conclusion and Discussion \label{sec:CD}}

In this paper we have modelled ad hoc and sensor networks as random geometric graphs, and investigated the effects of stochastic wireless channel models onto local and global network observables such as the mean degree, full connectivity and the minimum network degree.
Through analysis we have argued that compared to deterministic wireless channels, Rayleigh fading typically improves the network mean degree when $\eta<d$, and deteriorates it when $\eta>d$.
We have also argued that at high node densities a network which has minimum degree $1$ but is not fully connected will most likely consist of a large $N-2$ cluster and an isolated \textit{pair} of nodes. 
To this end, we have shown analytically and confirmed numerically that the probability $\Pi$ of isolated pairs forming in ad hoc networks is much less for stochastic wireless channels than for deterministic ones.

The question then arises as to how to interpret such a result in physical systems where the path loss exponent $\eta$ encodes the shadow fading and multipath effects to average received $\textrm{SNR}$.
In our view, a high path loss exponent suggests a higher level of correlations between neighbouring nodes.
That is, in heavily cluttered or lossy environments, two nearby nodes will be more correlated with regards to their network topology i.e. they will typically connect to the same nodes, rather than in less cluttered environments.
While hardware behaviour may be affected by other unconsidered factors, our model and subsequent theoretical analysis successfully captures the main physical picture and therefore has significant implications on real-world network emulation and deployment.
Namely, the fact that network behaviour is highly dependent upon whether a stochastic or deterministic connection model is employed must be taken into account by network design engineers, for one can envisage the potential for significantly over or under specifying network parameters if the incorrect model is chosen.  
Such an error could lead to costly deployments, or worse yet, networks that frequently experience faults or failures.

\bibliographystyle{IEEEtran}
\bibliography{mybib}

\begin{thebibliography}{10}
\providecommand{\url}[1]{#1}
\csname url@samestyle\endcsname
\providecommand{\newblock}{\relax}
\providecommand{\bibinfo}[2]{#2}
\providecommand{\BIBentrySTDinterwordspacing}{\spaceskip=0pt\relax}
\providecommand{\BIBentryALTinterwordstretchfactor}{4}
\providecommand{\BIBentryALTinterwordspacing}{\spaceskip=\fontdimen2\font plus
\BIBentryALTinterwordstretchfactor\fontdimen3\font minus
  \fontdimen4\font\relax}
\providecommand{\BIBforeignlanguage}[2]{{%
\expandafter\ifx\csname l@#1\endcsname\relax
\typeout{** WARNING: IEEEtran.bst: No hyphenation pattern has been}%
\typeout{** loaded for the language `#1'. Using the pattern for}%
\typeout{** the default language instead.}%
\else
\language=\csname l@#1\endcsname
\fi
#2}}
\providecommand{\BIBdecl}{\relax}
\BIBdecl

\bibitem{cordeiro2011ad}
C.~D.~M. Cordeiro and D.~P. Agrawal, \emph{Ad hoc and sensor networks: theory
  and applications}.\hskip 1em plus 0.5em minus 0.4em\relax World Scientific,
  2011.

\bibitem{savazzi2013ultra}
S.~Savazzi, U.~Spagnolini, L.~Goratti, D.~Molteni, M.~Latva-aho, and M.~Nicoli,
  ``Ultra-wide band sensor networks in oil and gas explorations,''
  \emph{Communications Magazine, IEEE}, vol.~51, no.~4, pp. 150--160, 2013.

\bibitem{santi2005topology}
P.~Santi, ``Topology control in wireless ad hoc and sensor networks,''
  \emph{ACM Computing Surveys (CSUR)}, vol.~37, no.~2, pp. 164--194, 2005.

\bibitem{xue2004number}
F.~Xue and P.~R. Kumar, ``The number of neighbors needed for connectivity of
  wireless networks,'' \emph{Wireless networks}, vol.~10, no.~2, pp. 169--181,
  2004.

\bibitem{stoyan1995stochastic}
D.~Stoyan, W.~S. Kendall, J.~Mecke, and L.~Ruschendorf, \emph{Stochastic
  geometry and its applications}.\hskip 1em plus 0.5em minus 0.4em\relax Wiley
  Chichester, 1995, vol.~2.

\bibitem{penrose2003random}
M.~Penrose, \emph{Random geometric graphs}.\hskip 1em plus 0.5em minus
  0.4em\relax Oxford University Press Oxford, UK:, 2003, vol.~5.

\bibitem{albert2002statistical}
R.~Albert and A.-L. Barab{\'a}si, ``Statistical mechanics of complex
  networks,'' \emph{Reviews of modern physics}, vol.~74, no.~1, p.~47, 2002.

\bibitem{coon2012full}
J.~Coon, C.~Dettmann, and O.~Georgiou, ``Full connectivity: corners, edges and
  faces,'' \emph{Journal of Statistical Physics}, pp. 1--21, 2012.

\bibitem{bettstetter2005connectivity}
C.~Bettstetter and C.~Hartmann, ``Connectivity of wireless multihop networks in
  a shadow fading environment,'' \emph{Wireless Networks}, vol.~11, no.~5, pp.
  571--579, 2005.

\bibitem{miorandi2008impact}
D.~Miorandi, ``The impact of channel randomness on coverage and connectivity of
  ad hoc and sensor networks,'' \emph{Wireless Communications, IEEE
  Transactions on}, vol.~7, no.~3, pp. 1062--1072, 2008.

\bibitem{tse2005fundamentals}
D.~Tse and P.~Viswanath, \emph{Fundamentals of wireless communication}.\hskip
  1em plus 0.5em minus 0.4em\relax Cambridge University Press, 2005.

\bibitem{zhou2008connectivity}
X.~Zhou, S.~Durrani, and H.~M. Jones, ``Connectivity of ad hoc networks: Is
  fading good or bad?'' in \emph{Signal Processing and Communication Systems,
  2008. ICSPCS 2008. 2nd International Conference on}.\hskip 1em plus 0.5em
  minus 0.4em\relax IEEE, 2008, pp. 1--5.

\bibitem{penrose1999k}
M.~Penrose, ``On k-connectivity for a geometric random graph,'' \emph{Random
  Structures \& Algorithms}, vol.~15, pp. 145--164, 1999.

\bibitem{haenggi2009stochastic}
M.~Haenggi, J.~Andrews, F.~Baccelli, O.~Dousse, and M.~Franceschetti,
  ``Stochastic geometry and random graphs for the analysis and design of
  wireless networks,'' \emph{Selected Areas in Communications, IEEE Journal
  on}, vol.~27, pp. 1029--1046, 2009.

\bibitem{fall2008dtn}
K.~Fall and S.~Farrell, ``\uppercase{DTN}: an architectural retrospective,''
  \emph{Selected Areas in Communications, IEEE Journal on}, vol.~26, no.~5, pp.
  828--836, 2008.

\bibitem{georgiou2013k}
O.~Georgiou, C.~P. Dettmann, and J.~Coon, ``k-connectivity for confined random
  networks,'' \emph{Europhysics Letters}, vol. 103, p. 28006, 2013.

\bibitem{buldyrev2010catastrophic}
S.~Buldyrev, R.~Parshani, G.~Paul, H.~Stanley, and S.~Havlin, ``Catastrophic
  cascade of failures in interdependent networks,'' \emph{Nature}, vol. 464,
  pp. 1025--1028, 2010.

\bibitem{bettstetter2004connectivity}
C.~Bettstetter, ``On the connectivity of ad hoc networks,'' \emph{The computer
  journal}, vol.~47, no.~4, pp. 432--447, 2004.

\bibitem{durgin2002new}
G.~D. Durgin, T.~S. Rappaport, and D.~A. De~Wolf, ``New analytical models and
  probability density functions for fading in wireless communications,''
  \emph{Communications, IEEE Transactions on}, vol.~50, no.~6, pp. 1005--1015,
  2002.

\end{thebibliography}
\end{document}